\documentclass[fleqn,11pt]{wlscirep}

\usepackage{tikz}
\usetikzlibrary{positioning,fit,matrix}
\usetikzlibrary{shapes.arrows,chains}
\graphicspath{{..}}

\usepackage{standalone}
\usepackage{helvet}
\usepackage{bm}
\usepackage{sansmath}
\usepackage{subcaption}
\usepackage{newtxtext,newtxmath}

\usepackage{afterpage}
\usepackage{placeins}

\usepackage{tikz}
\usetikzlibrary{positioning,fit}
\usetikzlibrary{matrix}
\usetikzlibrary{calc}
\usepackage{ifthen}
\usepackage{pgf}
\usepackage{pgffor}
\usepgfmodule{shapes}
\usepgfmodule{plot}
\usetikzlibrary{decorations}
\usetikzlibrary{arrows}
\usetikzlibrary{snakes}
\usepackage{pgfplots}
\usepackage{standalone}

\definecolor{darkgreen}{rgb}{0.0, 0.2, 0.13}
\definecolor{darkolivegreen}{rgb}{0.33, 0.42, 0.18}

\newcommand{\ie}{\textit{i.e.}\ }

\title{Modelling zebrafish collective behaviours with multilayer perceptrons optimised by evolutionary algorithms
}

\author[1,2]{Leo Cazenille}
\author[2]{Nicolas Bredeche}
\author[1,*]{José Halloy}

\affil[1]{Univ Paris Diderot, Sorbonne Paris Cit\'e, LIED, UMR 8236, 75013, Paris, France}
\affil[2]{Sorbonne Universit\'e, CNRS, ISIR, F-75005 Paris, France}

\affil[*]{jose.halloy@univ-paris-diderot.fr}


\begin{abstract}
Collective movements are pervasive behaviours among social organisms and have led to the development of many models. However, modelling animal trajectories and social interactions in simple bounded environments remains a challenge. Moreover, advances in the understanding of the sensory-motor loop and the information processing by animals are leading to revisions of the traditional assumptions made in decision-making algorithms.
In this context, we develop a methodology based on artificial neural networks (ANN) to describe the collective motion of small zebrafish groups in a bounded environment. Although ANN models are commonly used in artificial systems they are still under-explored to model animal collective behaviours. Here, we present a methodology to calibrate Multilayer Perceptrons by learning from real fish experimental data. 
The ANNs are trained using either supervised learning or various forms of evolutionary reinforcement learning methods (using the CMA-ES and NSGA-III algorithms). We reveal that ANN models trained using evolutionary methods are capable of generating realistic collective motions for groups of 5 zebrafish including the tank wall effects, a feature that is lacking in previous models. Finally, we also discuss the benefits of optimised ANNs as candidates for driving robotic lure with biologically realistic behaviour, a method that is becoming increasingly popular to gather data and validate assumptions on collective behaviours.

\end{abstract}
\begin{document}

\flushbottom
\maketitle
\thispagestyle{empty}

\section*{Introduction}

Collective motion is one of the most ubiquitous collective behaviour displayed by fish and can lead to collective decision-making~\cite{lopez2012behavioural}. An interest to study the link between individual behaviours and collective patterns has arisen out of these numerous observations and led to the development of models simulating agents performing collective movement and behaviours. A large variety of models exist to describe collective motion in biological systems~\cite{deutsch2012collective}. 

Modelling the collective motion of fish groups remains a challenging task, in particular for generating accurate collective trajectories\cite{herbert2015turing}. The first models were computer simulations describing the motion of a collection of individuals taking into account a collision avoidance and speed alignment mechanisms~\cite{Reynolds1987}. In this kind of simulations, the time step updating can be synchronous~\cite{Aoki1982,Aoki1984} or asynchronous~\cite{Bodeetal.2010,Bodeetal.2011}. Recent updated version of this kind of modelling has also been used~\cite{couzin2002collective,Lopezetal.2012}. In physics, models of self-propelled particles (SPP) have been developed to analyse the type of transitions between different group patterns from a phase transition point of view~\cite{Vicseketal.1995,Bertinetal.2006,Chateetal.2008,Chateetal.2010,Nagaietal.2015}. Inspired by mathematical models of random walk, kinematic models modelling fish trajectories with stochastic differential equations have also being developed to describe animal collective motions~\cite{Gautraisetal.2009, Gautraisetal.2012,Zienkiewiczetal.2014,Mwaffoetal.2014}. Most of these models consider animals as simple particles submitted to some kind of "social" forces that allow the emergence of collective motion patterns.

Probabilistic models based on a mechanism that takes into account visual perception and motion direction have been proposed~\cite{Lemassonetal.2009,Lemassonetal.2013,collignon2016stochastic}. The model considers the three dimensional visual sensory system of fish that adjust their trajectory according to their perception field. It introduces a stochastic process based on a probability distribution function to move in targeted directions rather than on a summation of influential vectors as it is assumed by previous models.

If the considered behaviours are more complex than aligned collective motion in unbounded environments more elaborate models are needed. Modelling realistic bounded and structured environments requires multilevel models that take into account perception of the environment and the other individuals and that allow decision-making~\cite{calovi2018disentangling,jiang2017identifying,lecheval2017domino}.
It is also an important issue to calibrate multilevel and spatially dependent social behaviour model because it involves trade-offs between social tendencies (\textit{e.g.} aggregation, group splitting), and response to the environment (\textit{e.g.} wall-following, zone occupation). We have previously shown that an optimisation method based on multi-objective evolutionary computation gives good results for complex models of collective behaviour~\cite{cazenille2015multi,cazenille2017LM,cazenille2016automated}.

\subsection*{Objectives}

Zebrafish is a classic animal model in genetics and neurosciences. It has been used extensively to study individual and collective behaviours, through experimental observations and modelling. Our goal is to automatically build models able to generate zebrafish trajectories that capture social interactions of individual fish within a small group of 5 fish placed in a bounded environment (i.e. with walls). We aim at generating models for autonomous fish agents that perform realistic collective behaviours for lasting duration of at least 30 minutes. This work participates to another general objective, namely to design robotic fish that can socially interact with live fish in realistic environments~\cite{romano2018review,cazenille2017acceptation}.

\subsection*{Contributions}

We consider groups of 5 agents including $n$ simulated autonomous agents driven by the MLP model and $(5-n)$ "real" agents with trajectories replayed from fish trajectories acquired from experimental observations. Varying the number $n$ of simulated agents makes it possible to tune the level of difficulty. A value of $n=1$ (easy setup) implies that the simulated agent should blend into a group performing realistic trajectories while a value of $n=5$ (difficult setup) requires that realistic trajectories are endogenous to the simulated agents.

Our method is summarised in Fig.~\ref{fig:workflow}. We rely on a multi-agents approach where each simulated agent is controlled by a Multilayer Perceptrons (MLP)~\cite{bishop2006prml} using high-level sensory inputs and actuators that are similar to what is available to zebrafish (see Table~\ref{tab:inputsOutputs}). MLP weights are trained using either supervised learning or stochastic optimisation techniques so as to generate relevant trajectories w.r.t. the expected behaviours. To do so, we define an objective function to maximise, which quantifies simulated agent individual trajectories (e.g. linear and angular speeds), group dynamics (e.g. inter-individual distances, polarisation) and relation with the environment (e.g. distance to wall). By using experimental data as a reference, the performance of MLP-driven simulated agents can be assessed.

Our results show that it is possible to design and automatically calibrate neural network models to act as simulated agent controllers. In particular, we reveal that evolutionary optimisation methods, namely CMA-ES~\cite{auger2005restart} and NSGA-III~\cite{yuan2014improved}, provide a great improvement over more classic supervised learning methods. This is due to the ability of evolutionary computation to act as a policy-search reinforcement learning method~\cite{sutton2018RL}, where sequential decisions are considered. This is very different than with supervised learning, where the goal is to devise effector outputs based solely on immediate sensory information without any regards from the sequence of actions (past or future). 

\section*{Methods} 

\subsection*{Animal handling and experimental set-up}
Our experiments involved 10 groups of 5 adults (6-12 months old) wild-type AB zebrafish (\textit{Danio rerio}) in ten 30-minutes trials as in~\cite{cazenille2016automated,seguret2017loose,cazenille2017acceptation,cazenille2018nn0,cazenille2018rtc}. Rearing and handling of the fish are described in the supplementary file. We use the experimental set-up that consists of a walled arena of $1\times1\times0.1$~m described in ~\cite{cazenille2016automated,seguret2017loose,cazenille2017acceptation,cazenille2018nn0,cazenille2018rtc} and detailed in the supplementary.

\begin{figure*}[h]
\begin{center}
\includegraphics[width=0.70\textwidth]{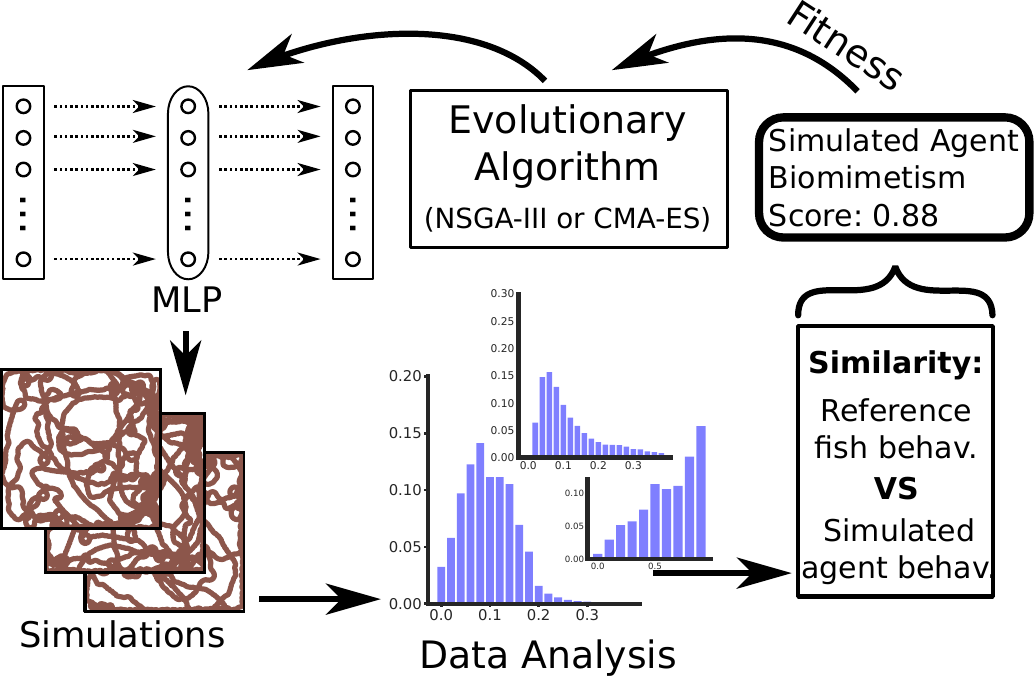}
\caption{Workflow of the methodology to optimise artificial neural networks to model fish behaviour. An evolutionary algorithm is used to evolve the weight of Multilayer Perceptron (MLP, 1 hidden layer, 100 neurons) that serve as behavioural models of fish-like agents in multi-agents simulations.  The fitness function is computed through data-analysis of these simulations and represent the biomimetism metric that quantifies the realism of the simulated fish behaviour compared to the behaviour exhibited by the experimental fish. Two evolutionary algorithms are tested: CMA-ES (mono-objective) and NSGA-III (multi-objective).}
\label{fig:workflow}
\end{center}
\end{figure*}

\subsection*{Artificial neural network model} \label{sec:model}
Black-box models, like artificial neural networks (ANN), can model phenomena with few \textit{a priori} information. To our knowledge, they were only used to model fish collective behaviour in preliminary studies~\cite{cazenille2018nn0,iizuka2018learning}. We extend the methodology from~\cite{cazenille2018nn0} based on experimental data, and we show that it is relevant to model zebrafish individual and collective behaviours.
This methodology (Fig.~\ref{fig:workflow}) makes use of Multilayer Perceptron (MLP)~\cite{bishop2006prml} artificial neural networks that are calibrated through the use of evolutionary algorithms to model the behaviour of simulated fish-like agents in a group of 5 individuals. Three cases with increasing difficulty are considered with groups composed of respectively 1,3,5 simulated agents driven by the calibrated MLP models and 4,2,0 agents following the experimental trajectories of experimental fish.

MLP are a type of feedforward artificial neural networks. They are very popular in artificial intelligence to solve a large variety of real-world problems~\cite{norgaard2000neural}. They have the capability to universally approximate functions~\cite{cybenko1989approximation} which makes them suitable to model control problems~\cite{norgaard2000neural}. We consider MLP with only one hidden layer of $100$ neurons (using a hyperbolic tangent function as activation function) as in~\cite{cazenille2018nn0}.

For the considered focal agent, the neural network model takes the $20$ parameters of Table~\ref{tab:inputsOutputs} as input. This set of inputs is typically used in multi-agent modelling of animal collective behaviour~\cite{deutsch2012collective,sumpter2012modelling}. We consider that it is sufficient to model fish behaviour with neural networks.

The neural network has $2$ outputs (Table~\ref{tab:inputsOutputs}) corresponding to the change in linear and angular speeds to apply from the current time-step to the next time-step. Our approach models fish trajectories resulting from social interactions in a homogeneous environment bounded by walls. Very few models of fish collective behaviours take into account the presence of walls~\cite{collignon2016stochastic,calovi2018disentangling,cazenille2018nn0,cazenille2017acceptation}.

\begin{table}[h]
\centering
\footnotesize

\begin{tikzpicture}
\node[inner sep=0] (inputs) at (0,0) {
\begin{tabular}{@{} p{4.0cm} p{1.5cm} p{11.0cm} @{}}
\hline
Name & \# of param. & Description \\
\hline
Linear speed & 1 & Instant linear speed of the focal agent at the previous time-step\\
Angular speed & 1 & Instant angular speed of the focal agent at the previous time-step\\
Distance towards agents & 4 & Linear distance from the focal agent towards each other agent\\
Angle towards agents & 4 & Angular distance from the focal agent towards each other agent\\
Alignment (angle) & 4 & Angular distance between the focal agent heading and each other agent heading\\
Alignment (linear speed) & 4 & Difference of linear speed between the focal agent and each other agent linear speed\\
Distance to nearest wall & 1 & Linear distance from the focal agent towards the nearest wall\\
Angle towards nearest wall & 1 & Angular distance from the focal agent towards the nearest wall\\
\hline
\end{tabular}
};

\node[inner sep=0,below=10mm of inputs] (outputs) {
\begin{tabular}{@{} p{4.0cm} p{1.5cm} p{11.0cm} @{}}
\hline
Name & \# of param. & Description \\
\hline
Delta linear speed & 1 & Change of instant linear speed of the focal agent from the previous time-step\\
Delta angular speed & 1 & Change of instant angular speed of the focal agent from the previous time-step\\
\hline
\end{tabular}
};

\sffamily
\fontsize{12}{25.0}\selectfont
\node[color=black,above left=0mm and 7mm of inputs,shift={(2.00,0.00)}] {\textbf{Inputs}};
\node[color=black,above left=0mm and 7mm of outputs,shift={(2.30,0.00)}] {\textbf{Outputs}};
\end{tikzpicture}

\caption{List of the $20$ parameters used in inputs and of the $2$ parameters used as outputs of the neural network models of agent behaviour.}
\label{tab:inputsOutputs}
\end{table}

\subsection*{Data analysis} \label{sec:dataAnalysis}
Zebrafish behaviour is multi-level (\ie with different dynamics at the level of individual trajectories \textit{vs.} at the level of group behaviour) and multi-modal (\ie with several kind of exhibited behaviours)~\cite{cazenille2017automated,cazenille2017acceptation,seguret2017loose}. In a bounded environment, zebrafish group exhibit a trade-off between aggregation and wall-following behaviours. Groups tend to be short-lived with individuals continuously leaving and joining the groups~\cite{cazenille2017acceptation}.

We quantify zebrafish behaviour by analysing each trial $e$ (experiments or simulations) and use the tracked positions of agents to compute the following behavioural metrics:
\begin{description}
    \item[at the level of individual trajectories] the distributions of \textit{instant linear speeds} ($L_e$) and the distributions of \textit{instant angular speeds} ($A_e$). These metrics assess the biomimetism of agent individual trajectories.
    
    \item[at the level of group dynamics] the distribution of \textit{inter-individual distances} between agents ($D_e$) and the distribution of \textit{polarisation} of the agents in the group ($P_e$). 
    The inter-individual distance evaluates the aggregative behaviour of the fish groups.
    The polarisation of an agent group measures how aligned the agents in a group are, and is defined as the absolute value of the mean agent heading: $P = \frac{1}{N} \bigl\lvert  \sum^{N}_{i=1} u_i \bigr\rvert$ where $u_i$ is the unit direction of agent $i$ and $N=5$ is the number of agents~\cite{tunstrom2013collective}.
    
    \item[in relation with the environment] the spatial (2D) distribution of \textit{probability (density) of presence in the arena} ($E_e$) and the distribution of \textit{distances of agents to their nearest wall} ($W_e$). These metrics assess the spatial repartition of the fish in arena.
\end{description}

We compute a \textbf{biomimetism score} using the similarity measure defined in~\cite{cazenille2018nn0,cazenille2017acceptation,cazenille2017automated} to quantify the realism of the behaviour of a group of simulated agents. It compares the behaviours of this group with the behaviours of the experimental fish averaged across all 10 experimental trials (\textbf{Control} case $e_c$). This score ranges from $0.0$ to $1.0$ and is defined as:
\begin{equation}
S(e, e_c) = \sqrt[5]{I(L_{e}, L_{e_c}) I(A_{e}, A_{e_c}) I(D_{e}, D_{e_c}) I(P_{e}, P_{e_c}) I(E_{e}, E_{e_c})}
\end{equation}
The function $I(X, Y)$ is defined as such: $I(X, Y) = 1 - H(X, Y)$.
The $H(X, Y)$ function is the Hellinger distance between two histograms~\cite{deza2006dictionary}. It is defined as: $H(X, Y) = \frac{1}{\sqrt{2}} \sqrt{ \sum_{i=1}^{d} (\sqrt{X_i} - \sqrt{Y_i}  )^2 }$ where $X_i$ and $Y_i$ are the bin frequencies.

\subsection*{Optimisation} \label{sec:optim}
We calibrate the MLP models describing agent behaviour to match as close as possible the dynamics of fish in groups of 5 individuals during 30 minutes simulations ($15$ time-steps per seconds, \ie $27000$ steps per simulation) using a methodology inspired from~\cite{cazenille2018nn0,cazenille2017automated}. This is achieved by optimising the connection weights of the MLP using evolutionary algorithms that iteratively perform global optimisation (inspired by biological evolution) on a defined fitness function so as to find its optimal value~\cite{salimans2017evolution,jiang2008supervised}. 

This calibration is a challenging problem because controllers have to cope with the multi-level and multi-modal nature of fish collective behaviour. It has to take into account conflicting behaviours and dynamics: dynamics at the level of the individual (individual trajectories), at the level of the group (aggregation) and in response to environmental stimuli (wall following behaviour). The optimisation process must handle an ill-defined fitness function (incorporating the biomimetic score defined earlier) that only describes partially the behaviour of experimental fish and that follow an aggregated representation (the biomimetic score aggregates several unrelated metrics).

We consider five optimisation methods (Table~\ref{tab:listCases}). Each case corresponds to a different way of optimising the MLP controllers and caters to different difficulties of the calibration process.
We employ two global optimisers: CMA-ES~\cite{auger2005restart} and NSGA-III. The CMA-ES~\cite{auger2005restart} algorithm is a popular state-of-the-art global optimiser that is able to handle problems where few assumptions need to be made on the nature of the underlying fitness function. It is able to cope with noisy, ill-defined fitness functions. It is a mono-objective optimisation algorithm (\ie the fitness function is unidimensional, only one objective is optimised). 
The NSGA-III~\cite{yuan2014improved} algorithm is a popular multi-objective (\ie the fitness function is multi-dimensional, several objectives are optimised) global optimiser . It is considered instead of the NSGA-II algorithm~\cite{deb2002fast} employed in~\cite{cazenille2017automated} because it is known to converge faster than NSGA-II on problems with more than two objectives~\cite{ishibuchi2016performance}.

In the \textbf{CMA-ES} case, we use the CMA-ES~\cite{auger2005restart} algorithm with the task of maximising the biomimetism score between MLP-driven agents simulations ($e_o$) and experiment fish groups ($e_c$): $S_{e_o, e_c}$. This case tests the hypothesis that a mono-objective solution is sufficient to cope with the multi-modal nature of the problem.

In the \textbf{NSGA-III-SP} (Selection pressures) case, we use the NSGA-III~\cite{yuan2014improved} to maximise three objectives. The first objective is a performance objective corresponding to the biomimetism score function: $S_{e_o, e_c}$. We also consider two other objectives used to guide the evolutionary process: one that promotes genotypic diversity~\cite{mouret2012encouraging} (defined by the mean euclidean distance of the genome of an individual to the genomes of the other individuals of the current population), the other encouraging behavioural diversity (defined by the euclidean distance between the $L_{e}$, $A_{e}$, $D_{e}$, $P_{e}$ and $E_{e}$ scores of an individual).
This case tests the hypothesis that the optimisation process needs a mechanism adapting the exploration vs exploitation tradeoff and encourage the exploration of solution with varying behaviours.

In the \textbf{NSGA-III-CM} (Combined Metrics) case, we use the NSGA-III~\cite{yuan2014improved} to maximise three objectives. The first objective is a performance objective corresponding to the biomimetism score function: $S_{e_o, e_c}$. The second and third objectives correspond to the aggregation of behavioural metrics related either to group behaviours (mean of inter-individual distance score $D_e$ and polarisation score $P_e$) or related to individuals trajectories and response to environmental factors (mean of linear speeds score $L_e$ and density of presence in the arena score $E_e$). This case tests the hypothesis that the optimisation process must explore solutions handling dynamics at different scales.

In the \textbf{NSGA-III-AM} (All metrics) case, we use the NSGA-III~\cite{yuan2014improved} to maximise five objectives: one corresponding to the biomimetim score function $S_{e_o, e_c}$, and the four others corresponding to the four behavioural distances metrics considered: the linear speeds score $L_e$, the inter-individual distance score $D_e$, the polarisation score $P_e$ and the density of presence score $E_e$. This case tests the hypothesis that the optimisation process must explore solutions catering to each considered behavioural metrics.

In the \textbf{NSGA-III-Nov} (Novelty) case, we use the NSGA-III~\cite{yuan2014improved} to maximise two objectives. The first objective is a performance objective corresponding to the biomimetism score function: $S_{e_o, e_c}$. The second objective is a novelty measure as defined in~\cite{lehman2008exploiting,mouret2011novelty}. It promotes the search of novel behaviours rather than only the search of efficiency (performance). This process has two benefits over optimising only performance: it can circumvent the deceptiveness present in some problems and it allows the evolutionary process to be more open-ended. Here, our problem is potentially deceptive has it involves trade-offs between different expected behaviours; their combination produces a large number of local optima in the fitness function.  The novelty measure $\rho(i)$ of individual $i$ is computed from a growing archive of previously encountered solutions (containing genomes and performances of previously evaluated individuals). It corresponds to the mean behavioural distance between $i$ and its $k$ nearest neighbours in the archive: $\rho(x) = \frac{1}{k} \sum^{k}_{j=0} d(x, \mu_j)$, with $\mu_j$ the $j$-th nearest neighbour of $x$, and $d$ the euclidean distance. Our best performing results are obtained with $k=4$.

In the \textbf{SL} (Supervised Learning) case, we employ a supervised learning approach to train MLPs to match experimental fish behaviour. The learning process is achieved by using the Adam~\cite{kingma2014adam} global optimiser. During training, a dropout layer (probability of dropout: $0.3$) is added after the hidden layer. The training dataset corresponds to the reconstructed sets of inputs and outputs for each agent of the experimental fish trajectory at each time-step. The loss function is the mean squared error compared to the reference outputs for a given set of inputs.

In all cases, we use populations of 120 individuals and 500 generations. Each case is repeated in 10 different trials.
We use CMA-ES and NSGA-III implementations based on the DEAP python library~\cite{fortin2012deap}.

\afterpage{
\clearpage
\begin{figure}[h]
\begin{center}
\begin{tikzpicture}
\sffamily
\fontsize{20}{25.0}\selectfont
\node[inner sep=0] (traj) at (0,0) {\includegraphics[width=0.99\textwidth]{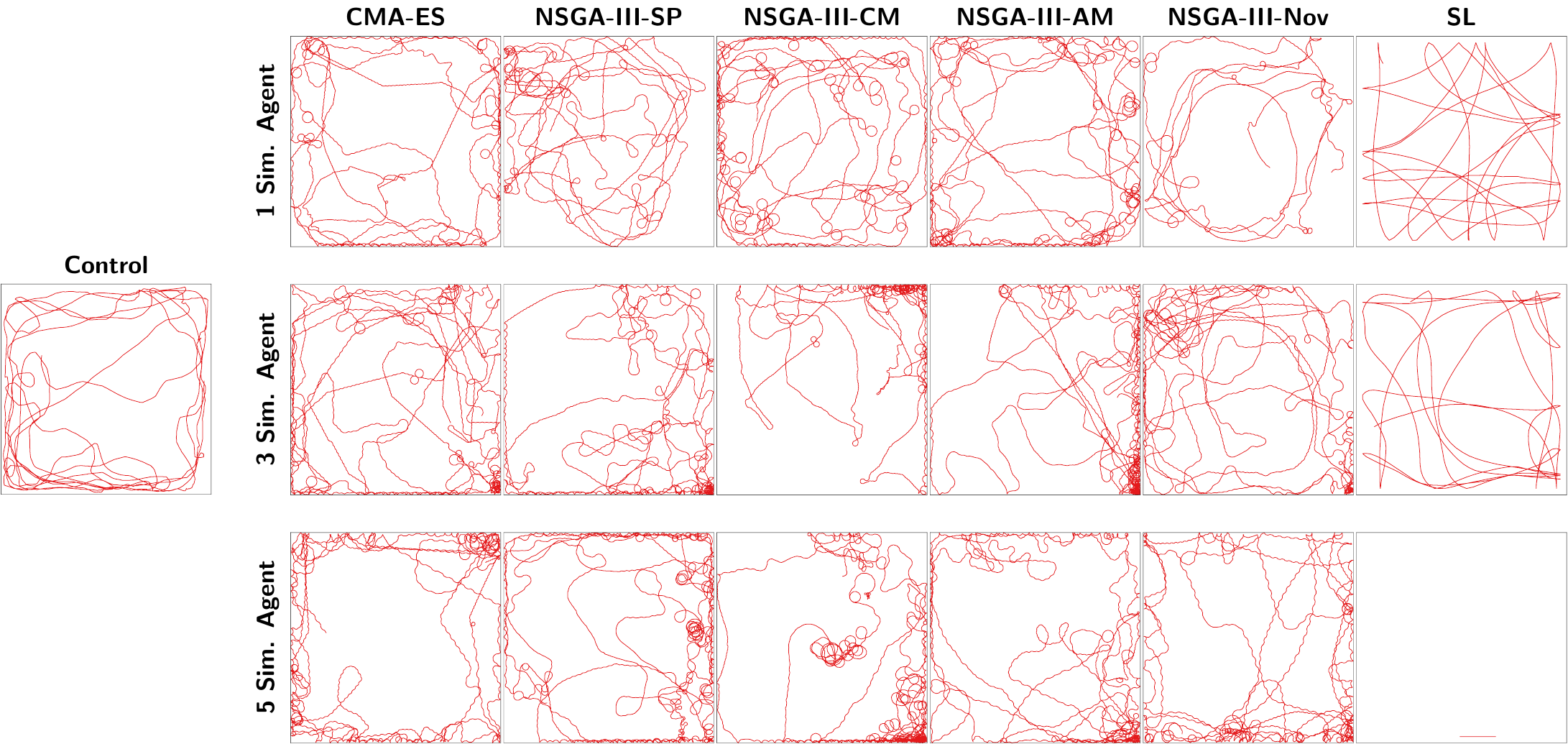}};
\node[inner sep=0,below=0mm of traj] (presences) {\includegraphics[width=0.99\textwidth]{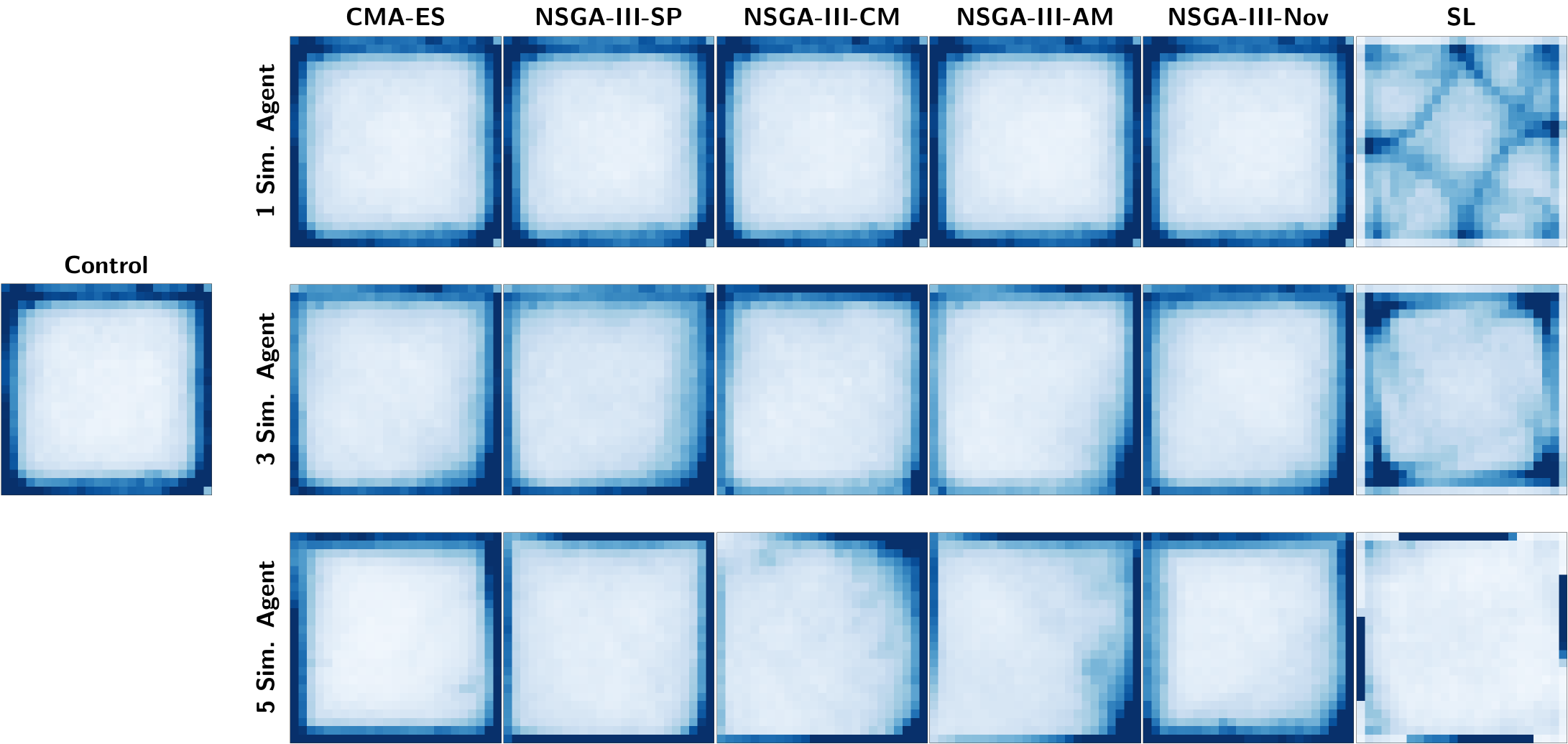}};
\node[color=black,above left=0mm and 5mm of traj,shift={(1.50,-2.00)}] {\textbf{A}};
\node[color=black,above left=0mm and 5mm of presences,shift={(1.50,-2.00)}] {\textbf{B}};
\end{tikzpicture}
\caption{Kinetic behaviour of the agents (simulated or experimental fish) in the arena, using the best-performing controllers. These controllers are MLP neural networks evolved through the five considered methods. They can be compared to  experimental fish (\textbf{Control} case) in 30-minutes trials in the square ($1m \times 1m$) arena, obtained as in~\cite{collignon2016stochastic,seguret2017loose}.  \textbf{A} Examples of individual trajectories of one simulated agent among the five making the group over one minute (900 time-steps). \textbf{B} Presence probability density of all autonomous agents in the arena represented by 2D histograms with $25\times25$ bins. }
\label{fig:plotsTracesPresences}
\end{center}
\end{figure}
\clearpage
}

\begin{table}[h]
\centering
\footnotesize
\begin{tabular}{@{} p{5cm} c c p{8cm} @{}}
\hline
Case name & Algorithm & \# of objectives & Objectives \\
\hline
\textbf{CMA-ES} & CMA-ES~\cite{auger2005restart} & 1 & Biomimetism score \\
\textbf{NSGA-III-SP} (Selection Pressures) & NSGA-III~\cite{yuan2014improved} & 3 & Biomimetism score, genotypic diversity, behavioural diversity \\
\textbf{NSGA-III-CM} (Combined Metrics) & NSGA-III~\cite{yuan2014improved} & 3 & Biomimetism score, trajectory metrics, environment metrics \\
\textbf{NSGA-III-AM} (All Metrics) & NSGA-III~\cite{yuan2014improved} & 5 & Biomimetism score, inter-individual score, linear speeds score, polarisation score, density of presence score \\
\textbf{NSGA-III-Nov} (Novelty) & NSGA-III~\cite{yuan2014improved} & 2 & Biomimetism score, novelty \\
\textbf{SL} & Adam~\cite{kingma2014adam} & - & Loss function: mean squared error \\
\hline
\end{tabular}
\caption{List of evolutionary algorithms methods used to optimise the MLP controllers weights. The best-performing controllers obtained through each method is compared to the experimental fish behaviours (\textbf{Control} case) of ten 30-minutes trials of five agents (simulated or experimental fish). The considered objectives are described in the Methods section. All evolutionary methods have a budget of 60 000 evaluations over 500 generations with populations of 120 individuals. Each methods is repeated through 10 different runs and fitness are computed across 10 different simulation trials per evaluation. The \textbf{CMA-ES} algorithm was used with an initial $\sigma=0.5$. The \textbf{NSGA-III} algorithm was used with rates of crossovers of $1.0$ and rates of mutations of $0.2$. In the \textbf{SL} case, MLP are trained by Adam~\cite{kingma2014adam} over $150$ epochs with a batch size of $64$ and a learning rate of $0.000005$. }
\label{tab:listCases}
\end{table}

\begin{figure}[h]
\centering
\includegraphics[width=0.80\textwidth]{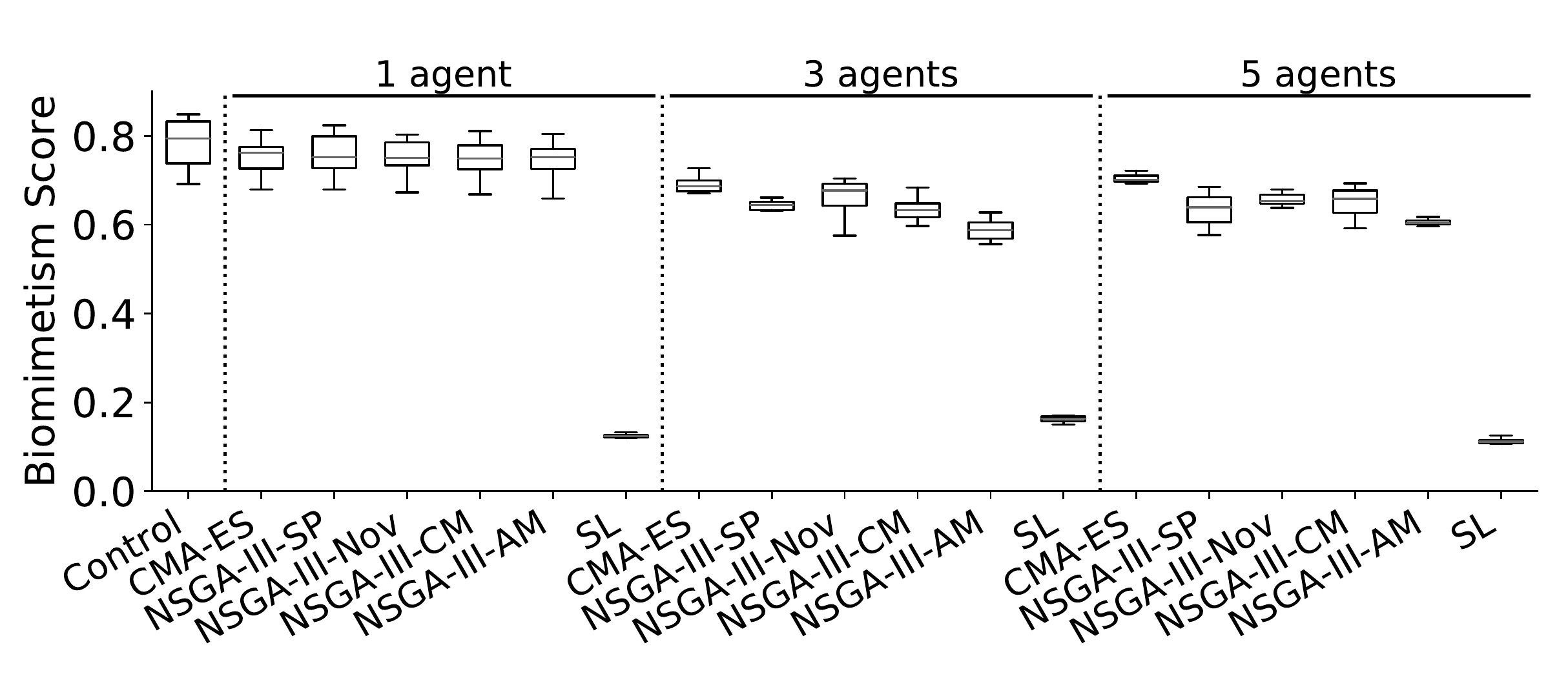}
\caption{Quantitative analysis of the difference of behaviour between the 5-agent experimental fish groups (\textbf{Control}) and 5-agent groups formed by 1,3,5 simulated agents and respectively 4,2,0 experimental fish from the \textbf{Control} case. The simulated agents are driven by the best-performing MLP models optimised by CMA-ES or NSGA-III algorithms and according to the configurations detailed on Table~\ref{tab:listCases}, obtained after 10 optimisation runs. Results are obtained over 10 different trials (experiments for fish-only groups, and simulations for ANN models). We consider five behavioural features to characterise exhibited behaviours. \textbf{Inter-individual distances} corresponds to the similarity in distribution of inter-individual distances between all agents and measures the capabilities of the agents to aggregate. \textbf{Linear speed distributions} corresponds to the distributions of linear speed of the agents. \textbf{Polarisation} measures how aligned the agents are in the group. \textbf{Probability of presence} corresponds to the similarity in spatial distribution of agent density of presence in each part of the arena. It involves the comparison of 2D spatial histograms as seen in Fig.~\ref{fig:plotsTracesPresences}B.  The \textbf{Biomimetic score} corresponds to the geometric mean of the other scores. Scores of simulated groups with 1 simulated agent optimised by evolutionary methods are not significantly different (Wilcoxon test, $p<0.05$) from experimental fish groups. However, scores of simulated groups with 3 or 5 simulated agents optimised by all methods are all significantly different from experimental fish groups.  }
\label{fig:scores}
\end{figure}

\section*{Results} \label{sec:results} 
We consider simulations with groups of 5 agents including 1, 3, or 5 simulated MLP-driven agents (Fig.~\ref{fig:workflow}) and respectively 4, 2, or 0 agents with trajectories corresponding to experimental fish data.
We analyse the behaviour the MLP-driven agents and quantify how they compare to the behaviour of fish-only groups (\textbf{Control} case). The MLP controllers are calibrated using the methods described in the previous section. We only consider the best-evolved ANN controllers.

Examples of agents trajectories are found in Fig.~\ref{fig:plotsTracesPresences}A. For all simulated cases, they correspond to the trajectories of an MLP-driven agents, and can be compared to the trajectories of an actual fish in the \textbf{Control} case. We can see that the MLP-driven agents exhibit wall-following behaviour as in \textbf{Control} but their trajectories incorporate patterns not found in actual fish trajectories. Namely, small circular loops can appear in all simulated cases (especially here with \textbf{NSGA-III} cases) when simulated agents performs U-turns to catch up with the rest of the group. Previous analyses in~\cite{cazenille2018nn0} showed that this effect cannot be easily mitigated even when the fitness function takes into account agents angular speeds.

We present the presence probability density of all agents in the arena in Fig.~\ref{fig:plotsTracesPresences}B. In experiments with only one MLP-driven agent, the resulting presence probability densities of all cases match those of the \textbf{Control} case. However, experiments with 3 or 5 MLP-driven agents can exhibit excesses of density of presence in corners compared to the \textbf{Control} case, despite that presence probability density is taken into account in the fitness function computation. It is especially present in experiments with 3 MLP-driven agents, and in the \textbf{NSGA-III-SP}, \textbf{NSGA-III-CM} and \textbf{NSGA-III-AM} cases.

We compute the biomimetism scores (Fig.~\ref{fig:scores}) presented in the Methods section quantifying the similarity between MLP-driven agents trajectories and \textbf{Control} case experimental fish trajectories. The similarity scores for each behavioural metric is found in Supplementary Information (Fig.~\ref{fig:scoresAll}).

The results of the \textbf{Control} case show a large variability in behaviour of the experimental fish because the 10 tested fish groups are composed of different fish (totalling 50 different fish) with different behaviour and individual preferences. This is especially the case with fish linear speeds (corroborating studies on zebrafish behaviour~\cite{seguret2017loose, collignon2017collective}) and wall-following behaviour (measured by the probability of presence scores). This suggests that fish follow walls from a distance that vary according to group composition, with sporadic (to a varying degree) passes through the centre of the arena, possibly in small short-lived sub-groups. Conversely, the fish tend to exhibit similar degree of alignment across the 10 experimental runs and sustained aggregative tendencies (measured by inter-individual distances scores).

All results obtained using the \textbf{SL} case are sub-optimal compared to results obtained through evolutionary computation.

The similarity scores of all evolutionary cases for groups with only 1 MLP-driven agent are within the variance domain of the \textbf{Control} case, except for the polarity score: it shows that these groups exhibit relatively similar dynamics as a fish-only group (at least according to our proposed measures). However, it remains difficult for the MLP-driven agent to stay aligned with the rest of the group. This is exemplified in Fig.~\ref{fig:plotsTracesPresences}A by the tendency of the MLP-driven agents to exhibit small trajectory loops, possibly as a sustained effort to remain close to the centroid of the group. Previous results in~\cite{cazenille2018nn0} showed that it is difficult to prevent this occurrence, even when agent angular speeds distributions are taken into account in fitness computation.

However, simulations of groups with 3 or 5 MLP-driven agents correspond to far more complex problem tasks and bring about results with mostly lower mean similarity scores than the \textbf{Control} case. In both types of groups, the \textbf{CMA-ES} case displays the best-performing individuals in term of biomimetic scores, with the \textbf{NSGA-III-AM} case corresponding to individuals with the lowest biomimetic scores amongst evolutionary cases. This is also the cases for inter-individual distances scores with results of the \textbf{CMA-ES}, \textbf{NSGA-III-Nov} and \textbf{NSGA-III-CM} cases with the variance domain of the \textbf{Control} case. All evolutionary cases display lower inter-individual distances scores for groups with 3 MLP-driven agents than with groups with 5 MLP-driven agents: it suggests that it is easier to aggregate with autonomous simulated agents capable of reacting to other agents (groups of 5 MLP-driven agents) rather than aggregate with the 2 non-responsive agents present in groups of 3 MLP-driven agents.

The situation is reversed with linear speeds scores, where most cases exhibit lower similarity scores with groups of 5 MLP-driven agents than in groups of 3 MLP-driven agents. Indeed, groups of 3 MLP-driven agents includes 2 agents with trajectories taken from experimental fish which would bias the scores to be closer to the \textbf{Control} case.

Depending on the behavioural metrics, different cases have varying degree of similarity to the \textbf{Control} case because they each have a different objective space, with several way of representing the target task. As such, some behaviours are easier to reach depending on the methodological case. With groups of 5 MLP-driven agents, the \textbf{CMA-ES} case has the highest scores of inter-individual distances, linear speeds and probability of presence. The \textbf{NSGA-III-CM} case has the highest polarity scores.

\section*{Discussion} 

Developing artificial neural network models for fish collective behaviours is a challenging issue because fish behaviours are multi-level (tail-beats as motor response vs individual trajectories vs collective dynamics), multi-modal (variety of behavioural patterns, input/output sources), context-dependent (behaviours depending on the spatial position and neighbours or group structure) and intrinsically stochastic (leading to individual and collectives choices and action selection)~\cite{collignon2016stochastic,sumpter2018using}. Moreover, fish dynamics involve trade-offs between social tendencies (aggregation, group formation), and response to the environment (wall-following, zone occupation). They also present distinct movement patterns that allow them to move in polarised groups and react collectively to environmental and social cues.

Here we show that evolved MLP models give good results for modelling zebrafish collective behaviours taking into account the boundaries of their environment. We quantify the realism of these MLP-driven autonomous agents compared to experimental fish by computing biomimetism scores.

Our methodology involves the calibration of these MLP models by using evolutionary computation to match experimental fish behaviour, using a biomimetism score as a performance objective. We employ and test both a mono-objective (CMA-ES~\cite{auger2005restart}) and a multi-objective (NSGA-III~\cite{yuan2014improved}) evolutionary algorithms to calibrate the weights of the MLP models. Here, we consider 1 to 5 autonomous MLP-driven agents. This allows to test models that do not need any fish present and still exhibit similar dynamics as fish groups. When considering more than one autonomous agent the task of optimising MLP models become far more difficult but it is more realistic as all modelled agents are capable of reacting to one another. 
We compare the results obtained through evolutionary computation to a supervised learning methodology where we train the MLP models on the reconstructed dataset of inputs and outputs for each time-step of the experimental data. This supervised learning method only provides sub-optimal results (similar to the results from~\cite{iizuka2018learning}), and are largely dominated by results obtained using evolutionary computation. The latter method differs mostly by the ability to pursue global optimisation, rather than local optimisation (supervised learning with gradient descent) which is prone to premature convergence.

Most of the models developed in behavioural biology describe only a small and specific part of the behavioural repertoire. Often, they do not completely describe animals as autonomous agents in their environment even in laboratory conditions. They are designed to answer specific biological questions. Moreover, the larger the behavioural repertoire, the larger the agent models, such that it becomes difficult to calibrate them on experimental data as the number of parameters explodes. The same issue arises to describe them by kinetic equations for the same reasons. In stark contrast, evolved ANN models can take into account behavioural details that are usually difficult to take into account in current behavioural models.

ANN models make it possible to ask relevant biological questions. Firstly, they can be used to find in the experimental data the necessary information needed to reproduce animal behaviours. They also open the door at using ANN to model the  functional neural working of the information processing by the animal during the sensory-motor loop. They do not need to be biomimetic at the lowest level \ie the individual neural physiology, but can model how information is processed. Finally, they open new possibilities to design models that can be transferred to robotics systems. Such system-based models allow to describe a fully autonomous agent in a realistic environment contrary to most models that describe only a sub-part of the agent behaviours~\cite{webb2009animals}.

Indeed, research on animal and robot interactions also need  bio-mimetic formal models as robot behavioural controllers for robots to behave like animals~\cite{cazenille2017acceptation,cazenille2017automated,bonnet2014miniature,Bonnet2016IJARS,bonnet2017cats}. Robots controllers have to deal with a whole range of behaviours to allow them to take into account not only other individuals but also environment particularities, such as  walls~\cite{cazenille2017acceptation,cazenille2017LM}. Robot behavioural models must be multilevel as they have to cope with low level hardware events and high level social behaviours. In this study, we showed that ANN are good candidates to model individual and collective fish behaviours, which is particularly relevant in the context of social bio-hybrid systems composed of animals and robots.

\FloatBarrier

\section*{Acknowledgements}

This work was funded by EU-ICT project ASSISIbf, $n^{o} 601074$. We thank the RER B for hosting the discussion leading to this study.

\section*{Author contributions statement}
L.C. designed the study and performed the data analysis, ran the simulations and wrote the manuscript. N.B. contributed with artificial neural networks and evolutionary computation expertise, wrote and revised the manuscript. J.H. designed and coordinated the study, wrote and revised the manuscript. All authors gave final approval for publication.

\section*{Additional information}

\textbf{Accession codes} (where applicable)

\textbf{Competing financial interests:} The authors declare no competing  financial interests.

\clearpage
\appendix
\section*{Supplementary Information}

\subsection*{Ethical statement} 
The experiments performed in this study were conducted under the authorisation of the Buffon Ethical Committee (registered to the French National Ethical Committee for Animal Experiments \#40) after submission to the French state ethical board for animal experiments.

\subsection*{Fish rearing and handling}

The fish were 6-12 months old at the time of the experiments. We kept the fish under laboratory conditions, 27$\ensuremath{^\circ}$C, 500$\mu$S salinity with a 10:14 day:night cycle. The fish were reared in housing facilities ZebTEC and fed two times a day (Special Diets Services SDS-400 Scientific Fish Food). The water pH level was maintained at 7, and Nitrites (NO$^{-2}$) were below $0.3$~mg/l.

\subsection*{Experimental set-up} \label{sec:setup}
We use the experimental set-up described in ~\cite{cazenille2016automated,seguret2017loose,cazenille2017acceptation,cazenille2018nn0,cazenille2018rtc}: it consists of an arena of $1000\times1000\times100$~mm made of white plexiglass and placed in a $1200\times1200\times300$~mm experimental tank.
The tank is filled with water up to a level of $60$~mm. The whole set-up is exposed to diffused light and confined behind white sheets to isolate experiments and homogenise luminosity.
We use an overhead acA2040-25gm monochrome GigE CCD camera (Basler AG, Germany) to grab video frames of the experiments (with a resolution of $2040\times2040$ pixels, $15$ frames per second). This camera is equipped with low distortion lenses CF12.5HA-1 (Fujinon, Tokyo, Japan). The experimental videos are then analysed and tracked off-line to retrieve the position of the fish~\cite{perez2014idtracker}.

\subsection*{Supplementary results}

Figure~\ref{fig:scoresAll} shows the behavioural scores of the best-performing controllers obtained through the studied optimisation methods.

Figures~\ref{fig:plotsHistsCMAES},\ref{fig:plotsHistsNSGAIIISP},\ref{fig:plotsHistsNSGAIIICM},\ref{fig:plotsHistsNSGAIIIAM},\ref{fig:plotsHistsNSGAIIINov},\ref{fig:plotsHistsSL} provide analyses of simulated fish groups behaviour obtained through different optimisation methods (Table~\ref{tab:listCases}): respectively \textbf{CMA-ES}, \textbf{NSGA-III-SP}, \textbf{NSGA-III-CM}, \textbf{NSGA-III-AM}, \textbf{NSGA-III-Nov} and \textbf{SL}. These figures take into account the following metrics:
\textbf{Inter-individual distances} corresponds to the similarity in distribution of inter-individual distances between all agents and measures the capabilities of the agents to aggregate. \textbf{Linear speed distributions} corresponds to the distributions of linear speed of the agents. \textbf{Polarisation} measures how aligned the agents are in the group. \textbf{Probability of presence} corresponds to the similarity in spatial distribution of agent density of presence in each part of the arena. It involves the comparison of 2D spatial histograms as seen in Fig.~\ref{fig:plotsTracesPresences}B.  The \textbf{Biomimetic score} corresponds to the geometric mean of the other scores. 

Figure~\ref{fig:fitnessPerEvals} shows the evolution of biomimetic score in fitness computation per evaluation.

\afterpage{

\begin{figure}[h]
\centering

\includegraphics[width=0.60\textwidth]{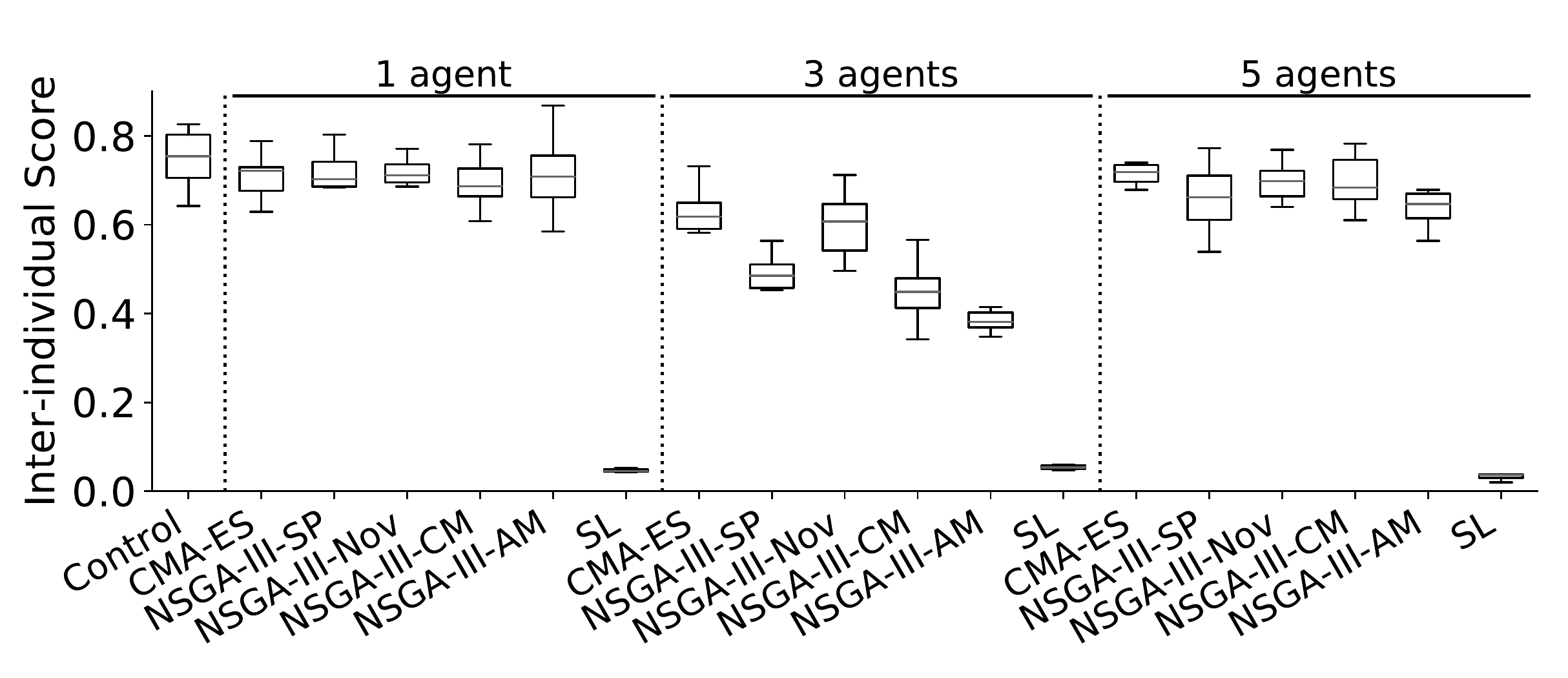}

\includegraphics[width=0.60\textwidth]{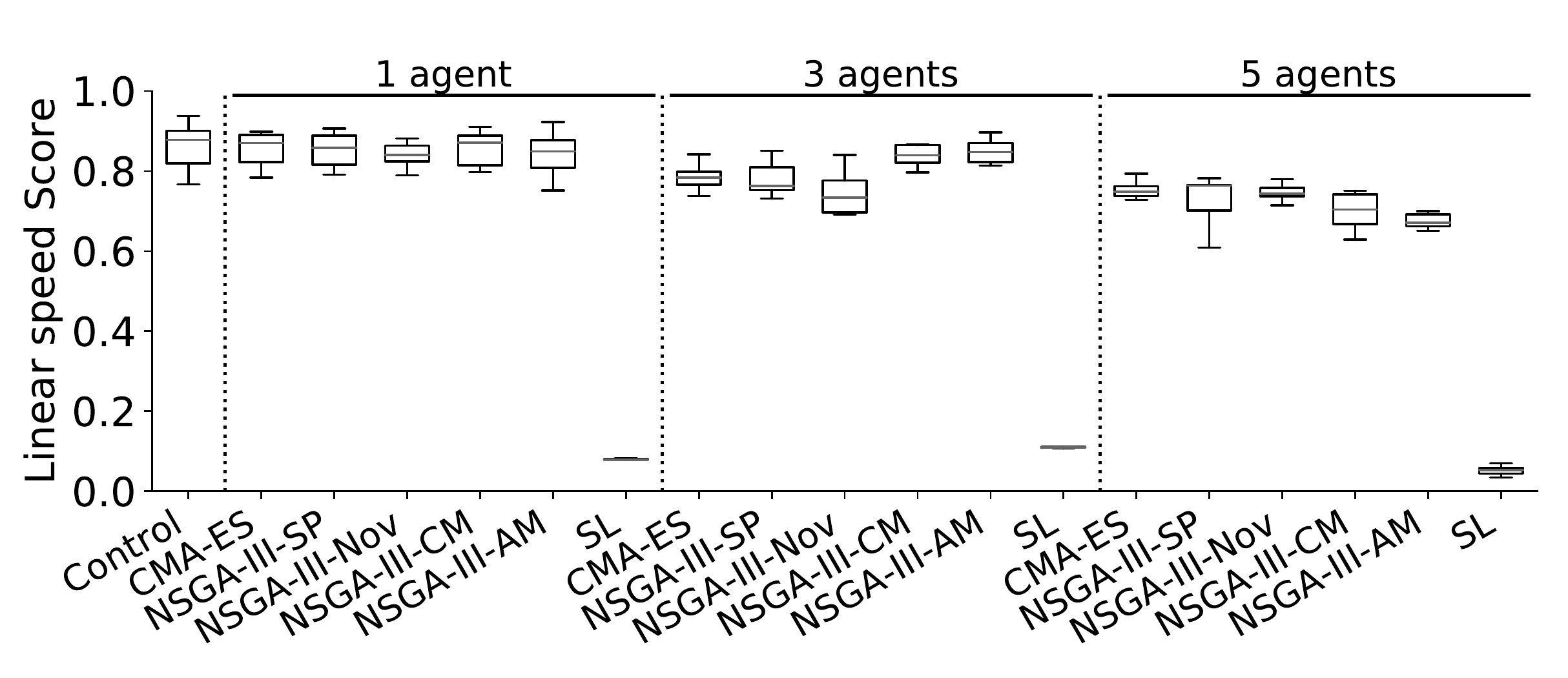}

\includegraphics[width=0.60\textwidth]{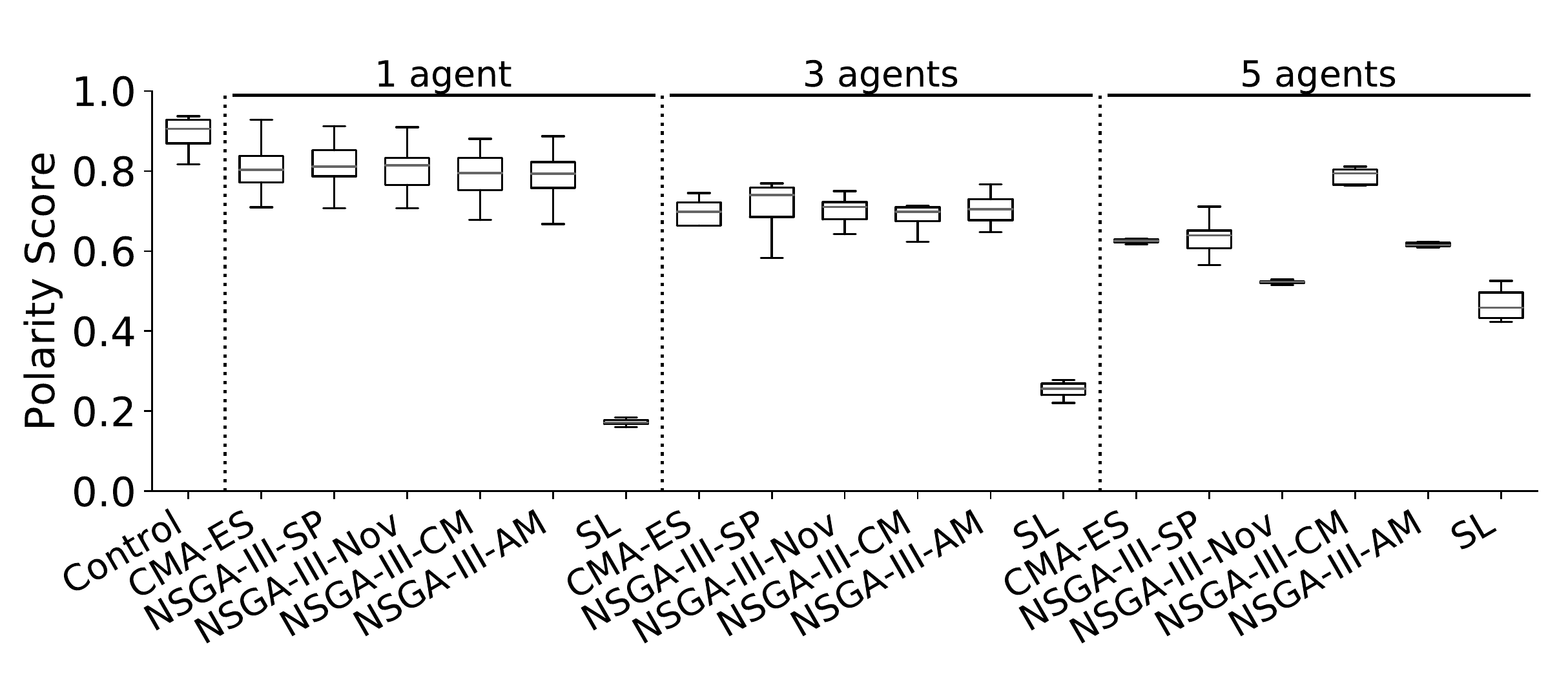}

\includegraphics[width=0.60\textwidth]{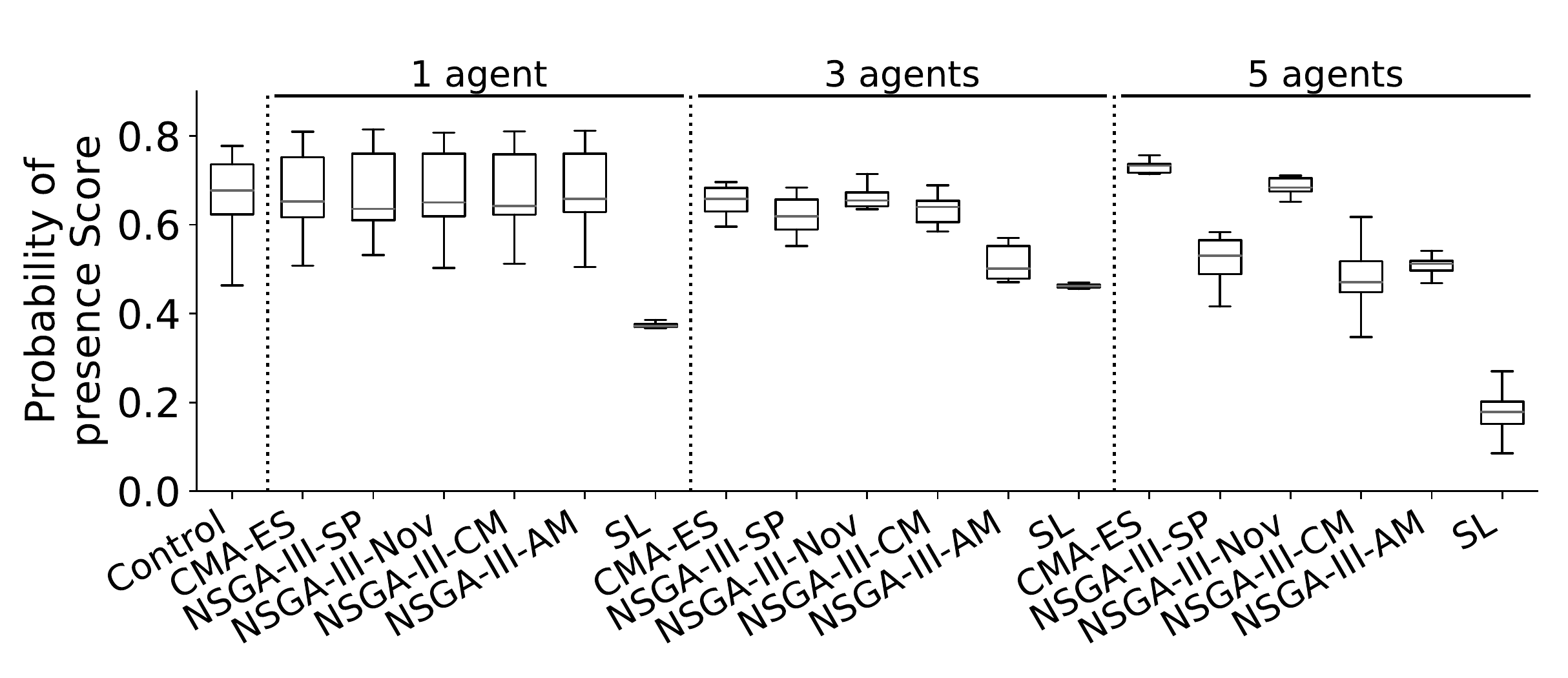}
\caption{Quantitative analysis of the difference of behaviour between the 5-agent experimental fish groups (\textbf{Control}) and 5-agent groups formed by 1,3,5 simulated agents and respectively 4,2,0 experimental fish from the \textbf{Control} case. The simulated agents are driven by the best-performing MLP models optimised by CMA-ES or NSGA-III algorithms and according to the configurations detailed on Table~\ref{tab:listCases}, obtained after 10 optimisation runs. Results are obtained over 10 different trials (experiments for fish-only groups, and simulations for ANN models). We consider five behavioural features to characterise exhibited behaviours. \textbf{Inter-individual distances} corresponds to the similarity in distribution of inter-individual distances between all agents and measures the capabilities of the agents to aggregate. \textbf{Linear speed distributions} corresponds to the distributions of linear speed of the agents. \textbf{Polarisation} measures how aligned the agents are in the group. \textbf{Probability of presence} corresponds to the similarity in spatial distribution of agent density of presence in each part of the arena. It involves the comparison of 2D spatial histograms as seen in Fig.~\ref{fig:plotsTracesPresences}B.  Similarity scores of simulated groups with 1 simulated agent optimised by all methods are not significantly different (Wilcoxon test, $p<0.05$) from experimental fish groups (except for the polarity metric). However, Similarity scores of simulated groups with 3 or 5 simulated agents optimised by all methods are all significantly different from experimental fish groups.  }
\label{fig:scoresAll}
\end{figure}

\clearpage
\begin{figure}[h]
\begin{center}
\includegraphics[width=0.90\textwidth]{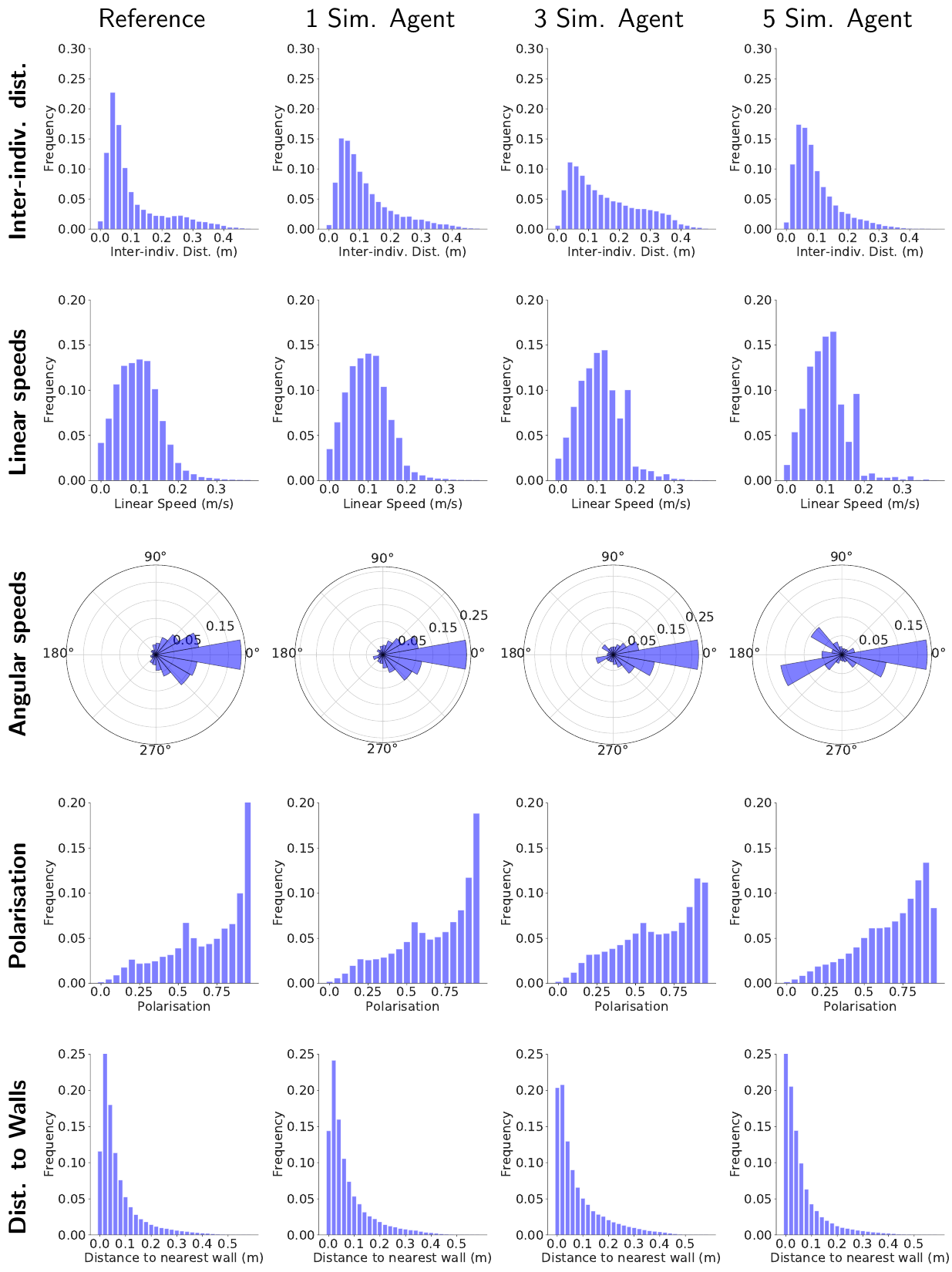}
\caption{Comparison of agents behaviours between the \textbf{Control} case (10 trials of 30-minutes experiments involving 5 zebrafish) and the \textbf{CMA-ES} case (10 trials of 30-minutes simulations of 5 agents, with respectively 1,3,5 simulated agents driven by the best-performing MLP controller optimised through the \textbf{CMA-ES} method and 4,2,0 agents copying the trajectories of experimental fish). Agents behaviour is quantified across 5 behavioural features: inter-individual distances, linear and angular speeds distributions, polarisation, and distances to nearest wall.}
\label{fig:plotsHistsCMAES}
\end{center}
\end{figure}
\clearpage

\begin{figure}[h]
\begin{center}
\includegraphics[width=0.90\textwidth]{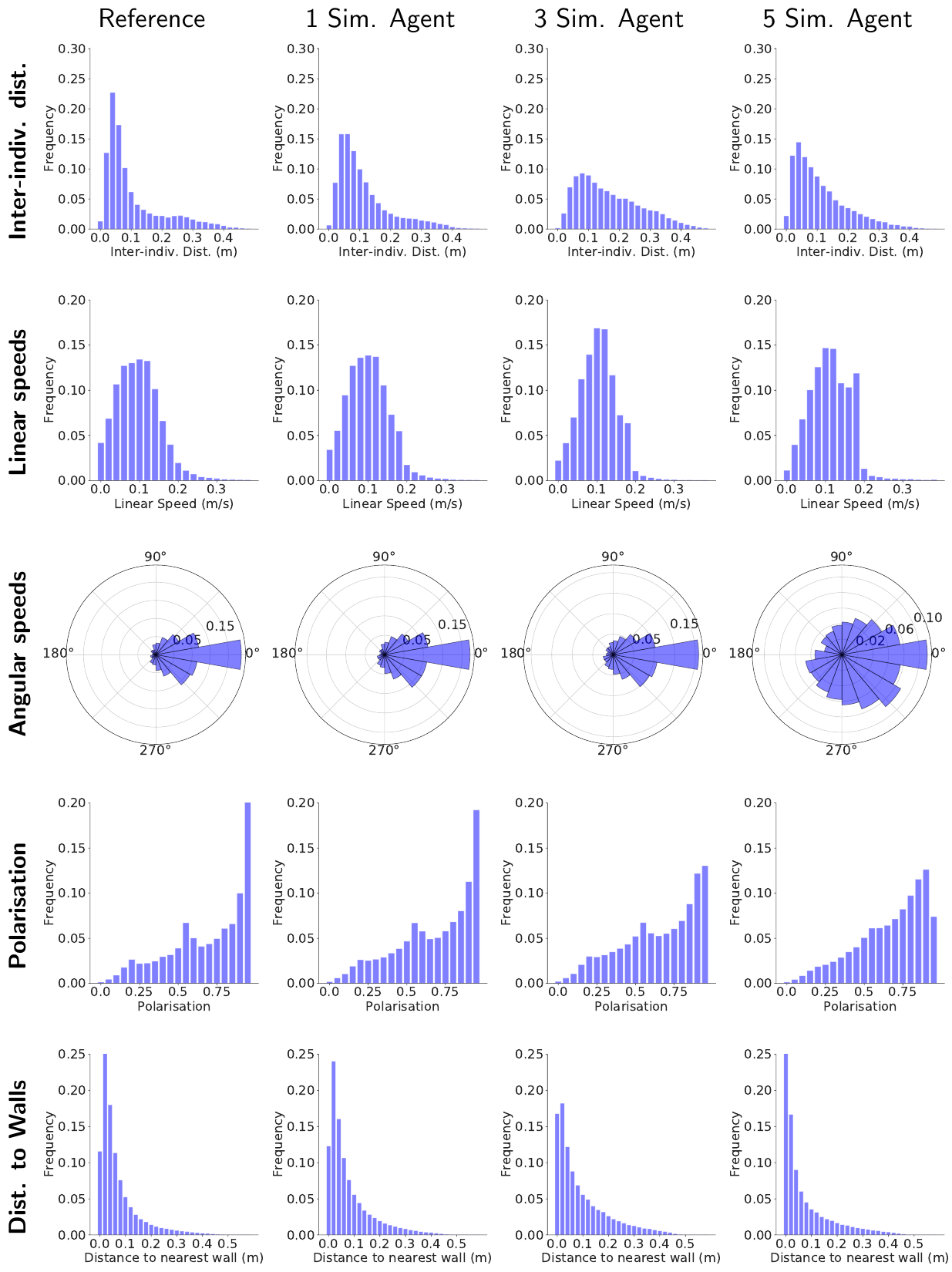}
\caption{Comparison of agents behaviours between the \textbf{Control} case (10 trials of 30-minutes experiments involving 5 zebrafish) and the \textbf{NSGA-III-SP} case (10 trials of 30-minutes simulations of 5 agents, with respectively 1,3,5 simulated agents driven by the best-performing MLP controller optimised through the \textbf{NSGA-III-SP} method and 4,2,0 agents copying the trajectories of experimental fish). Agents behaviour is quantified across 5 behavioural features: inter-individual distances, linear and angular speeds distributions, polarisation, and distances to nearest wall.}
\label{fig:plotsHistsNSGAIIISP}
\end{center}
\end{figure}
\clearpage

\begin{figure}[h]
\begin{center}
\includegraphics[width=0.90\textwidth]{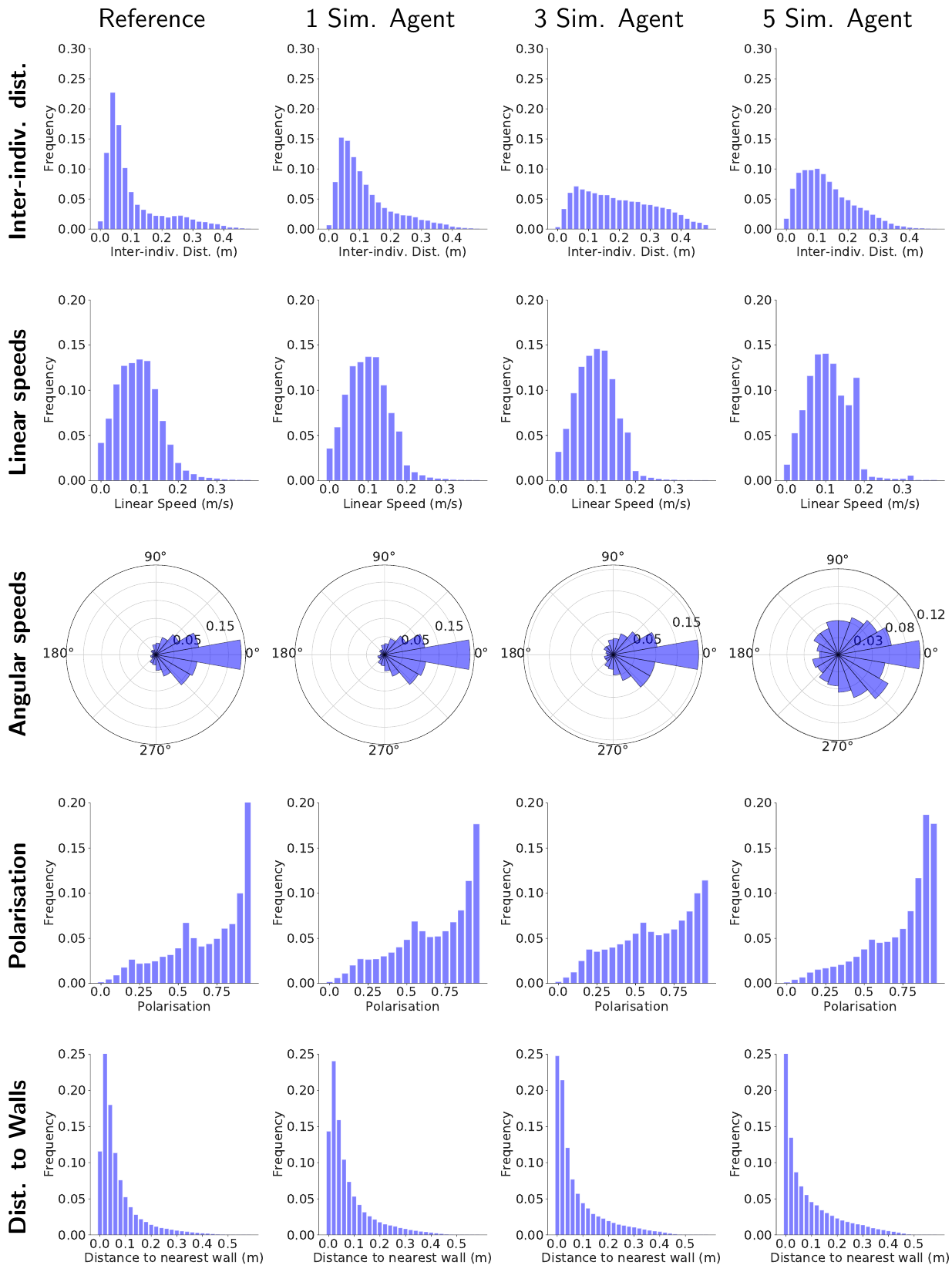}
\caption{Comparison of agents behaviours between the \textbf{Control} case (10 trials of 30-minutes experiments involving 5 zebrafish) and the \textbf{NSGA-III-CM} case (10 trials of 30-minutes simulations of 5 agents, with respectively 1,3,5 simulated agents driven by the best-performing MLP controller optimised through the \textbf{NSGA-III-CM} method and 4,2,0 agents copying the trajectories of experimental fish). Agents behaviour is quantified across 5 behavioural features: inter-individual distances, linear and angular speeds distributions, polarisation, and distances to nearest wall.}
\label{fig:plotsHistsNSGAIIICM}
\end{center}
\end{figure}
\clearpage

\begin{figure}[h]
\begin{center}
\includegraphics[width=0.90\textwidth]{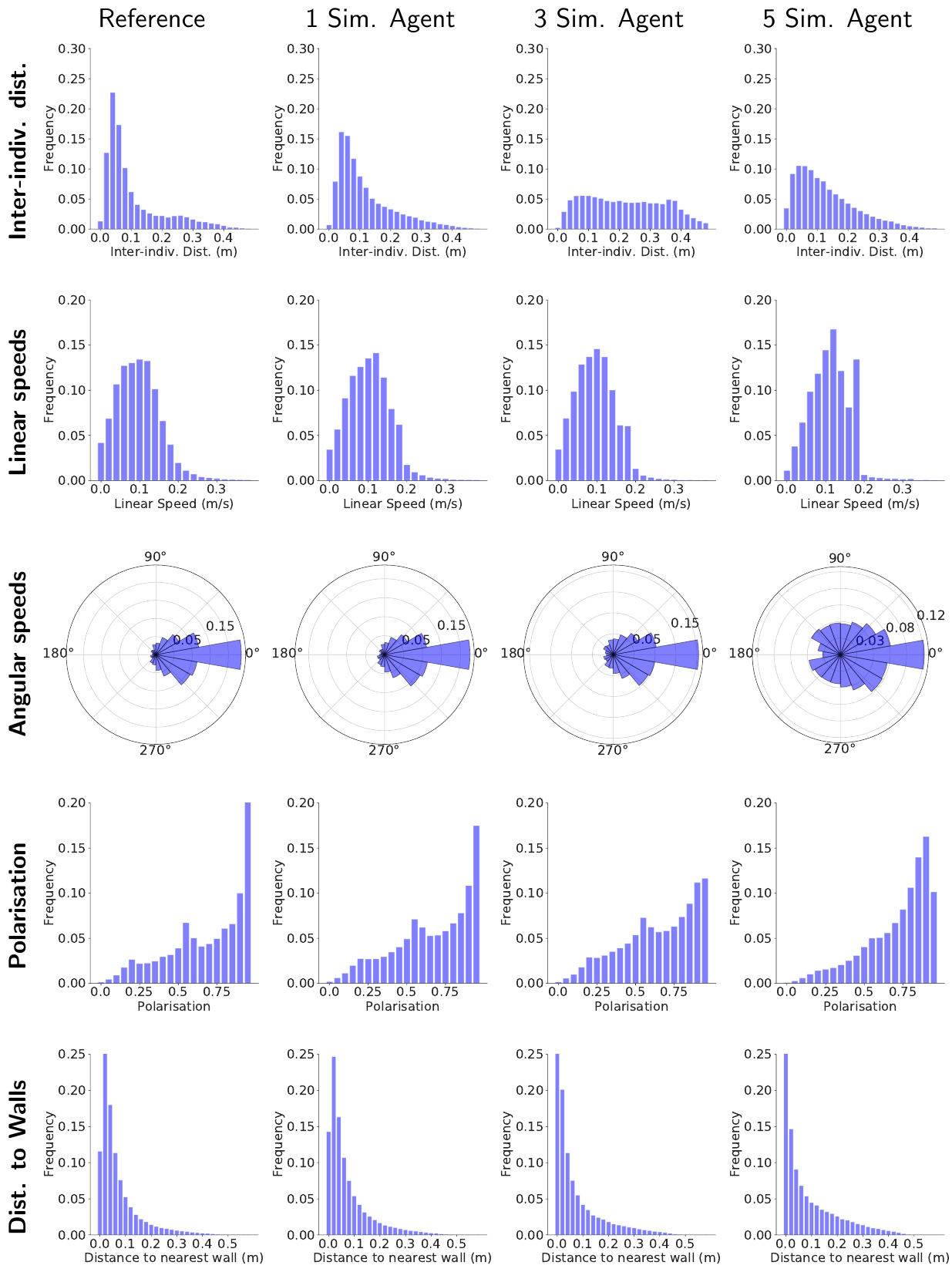}
\caption{Comparison of agents behaviours between the \textbf{Control} case (10 trials of 30-minutes experiments involving 5 zebrafish) and the \textbf{NSGA-III-AM} case (10 trials of 30-minutes simulations of 5 agents, with respectively 1,3,5 simulated agents driven by the best-performing MLP controller optimised through the \textbf{NSGA-III-AM} method and 4,2,0 agents copying the trajectories of experimental fish). Agents behaviour is quantified across 5 behavioural features: inter-individual distances, linear and angular speeds distributions, polarisation, and distances to nearest wall.}
\label{fig:plotsHistsNSGAIIIAM}
\end{center}
\end{figure}
\clearpage

\begin{figure}[h]
\begin{center}
\includegraphics[width=0.90\textwidth]{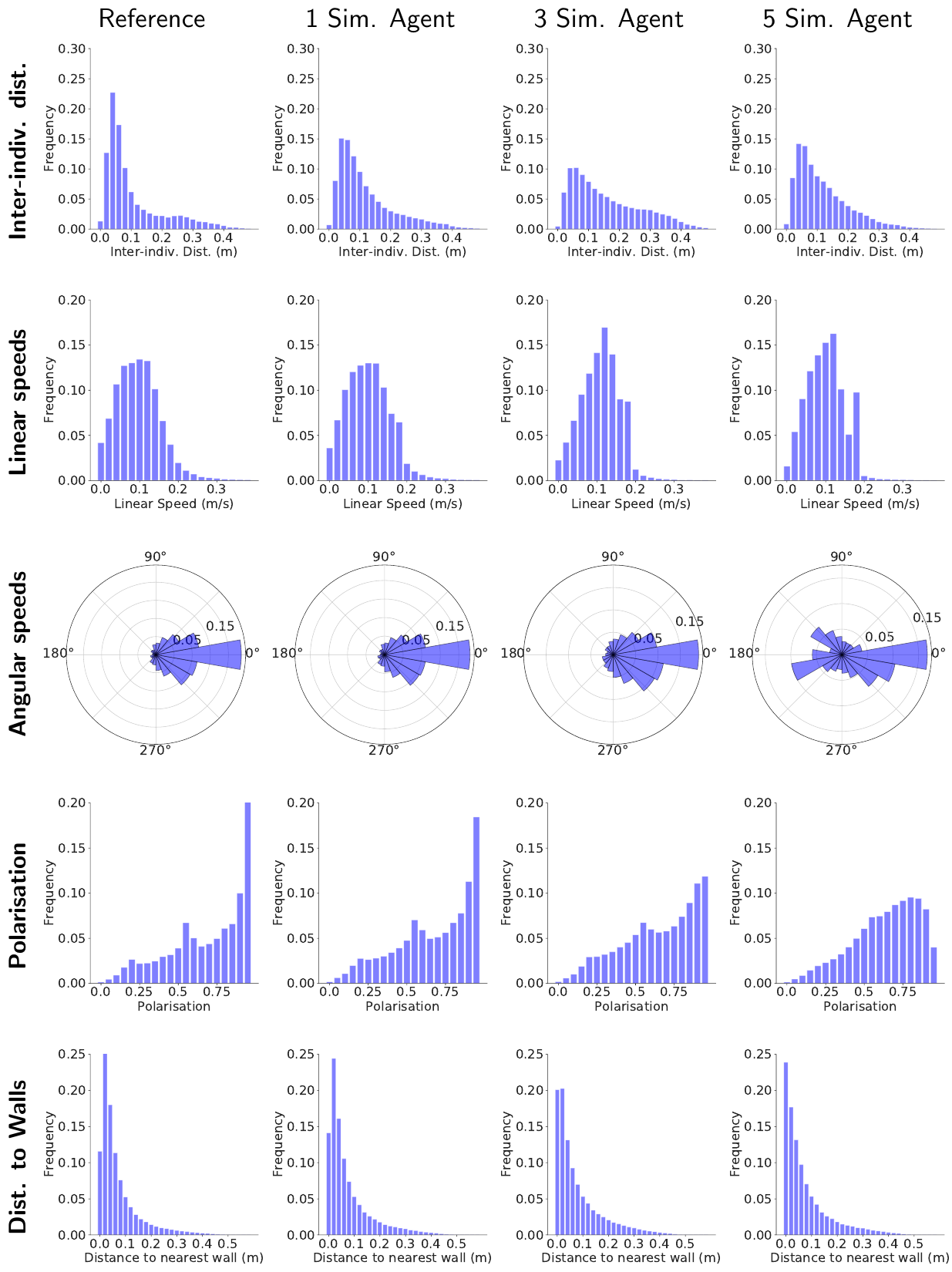}
\caption{Comparison of agents behaviours between the \textbf{Control} case (10 trials of 30-minutes experiments involving 5 zebrafish) and the \textbf{NSGA-III-Nov} case (10 trials of 30-minutes simulations of 5 agents, with respectively 1,3,5 simulated agents driven by the best-performing MLP controller optimised through the \textbf{NSGA-III-Nov} method and 4,2,0 agents copying the trajectories of experimental fish). Agents behaviour is quantified across 5 behavioural features: inter-individual distances, linear and angular speeds distributions, polarisation, and distances to nearest wall.}
\label{fig:plotsHistsNSGAIIINov}
\end{center}
\end{figure}
\clearpage

\begin{figure}[h]
\begin{center}
\includegraphics[width=0.90\textwidth]{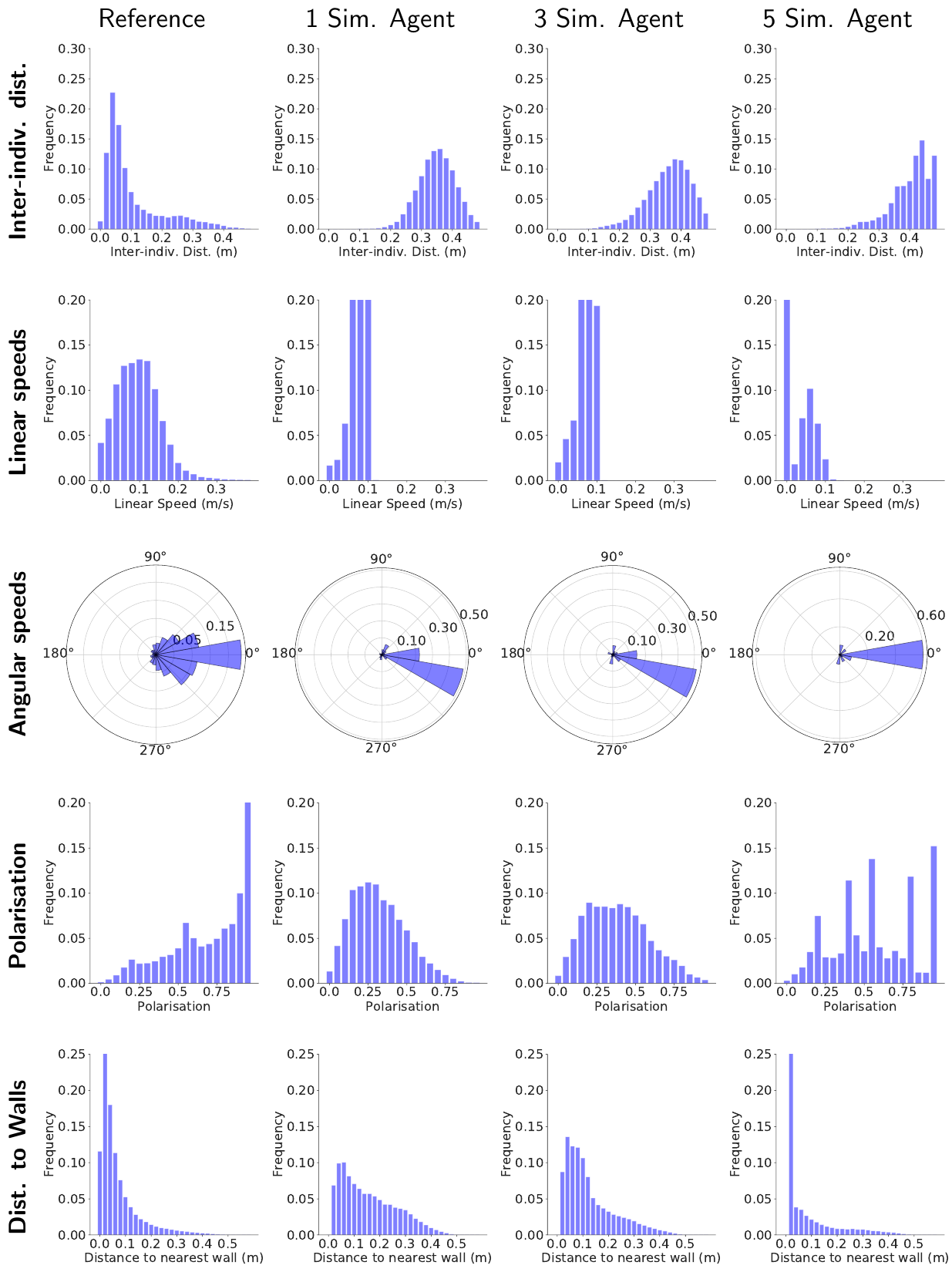}
\caption{Comparison of agents behaviours between the \textbf{Control} case (10 trials of 30-minutes experiments involving 5 zebrafish) and the \textbf{SL} case (10 trials of 30-minutes simulations of 5 agents, with respectively 1,3,5 simulated agents driven by the best-performing MLP controller optimised through the \textbf{SL} method and 4,2,0 agents copying the trajectories of experimental fish). Agents behaviour is quantified across 5 behavioural features: inter-individual distances, linear and angular speeds distributions, polarisation, and distances to nearest wall.}
\label{fig:plotsHistsSL}
\end{center}
\end{figure}
\clearpage

\begin{figure}[h]
\begin{center}
\includegraphics[width=0.99\textwidth]{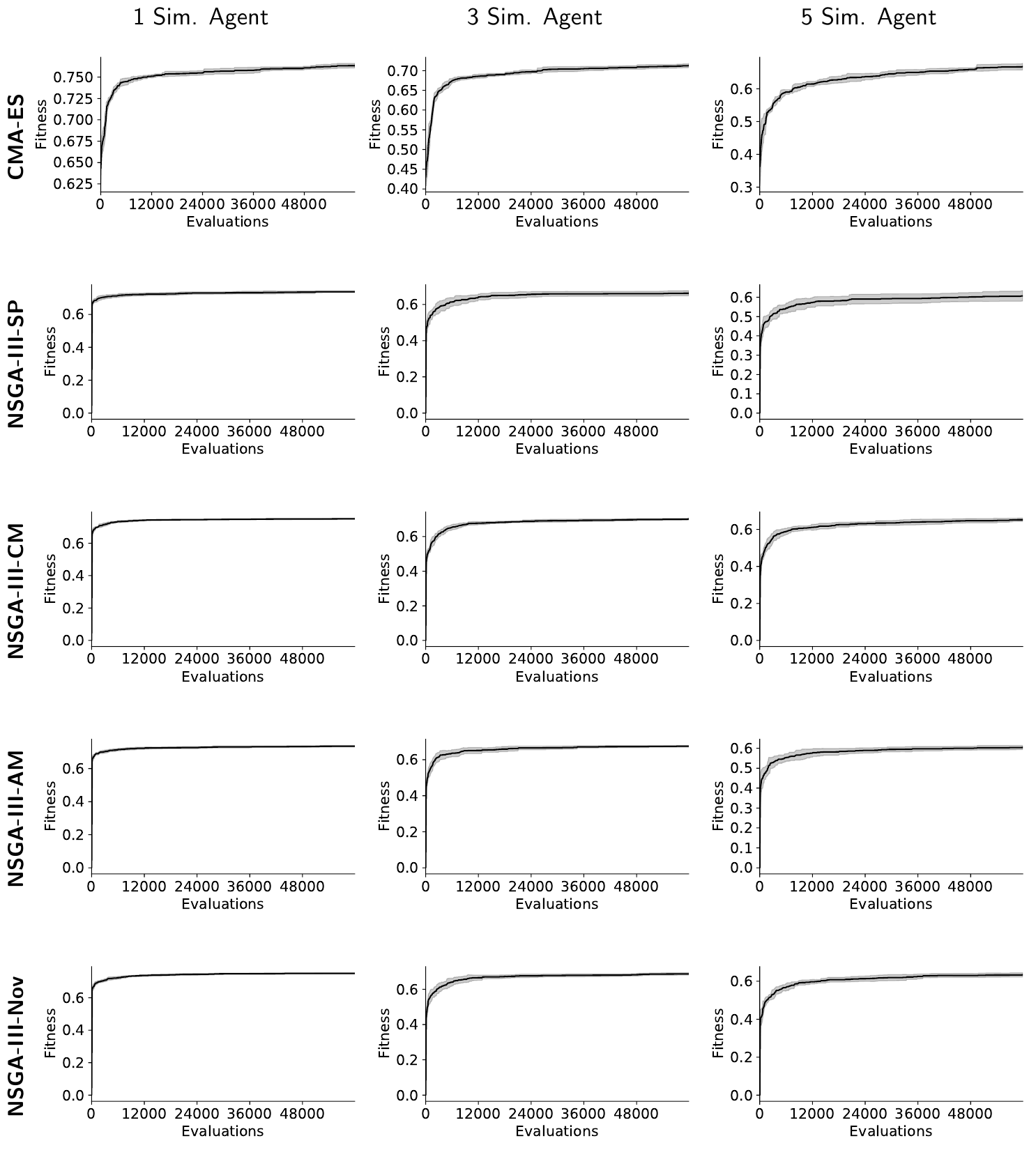}
\caption{Evolution of the biomimetism score used in the computation of the fitness values across evaluations for all tested optimisation method. Each optimisation method is tested in 10 different runs. Each evaluation involves 10 trials of 5-minutes simulations of 5 agents, with respectively 1,3,5 simulated agents driven by the tested MLP controllers (phenotypes of the evolution process) and 4,2,0 agents copying the trajectories of experimental fish. }
\label{fig:fitnessPerEvals}
\end{center}
\end{figure}
}

\end{document}